# Possible origin of the interface field in spintronic experiments

Oleg Tikhomirov

Institute of Solid State Physics RAS, Chernogolovka, Russia 142432

The effective field at the interface between ferromagnetic and metal layers is often observed in spintronic experiments. It is common to ascribe it to specific exchange interactions caused by spin-orbit coupling, namely, Rashba field, Dzyaloshinskiy-Moriya interaction, spin currents, etc. Alternative view involves the global magnetic field (Oersted field) generated by current both inside and outside the conductor. It is shown here that the inhomogeneous stray field surrounding ferromagnetic layers can produce significant contribution into the observed interface field. This mechanism should be taken into account for better understanding the current-induced phenomena in magnetic multilayers.

The progress in magnetism is often driven by a new concept of magnetic memory. Current boom in spintronics is seriously motivated by idea of the "race-track memory" where ferromagnetic domain walls proceed along separate wires to carry bits of information[1]. Owing to this project, both experimental and theoretical investigation of electric current with ferromagnetic domain structure is under way.

Beyond direct observations of domain walls before and after electric pulse using the Kerr effect and other microscopic techniques, indirect measurements via magnetoresistance or Hall effect are widely used. In many cases the measured effect is asymmetric with respect to additional magnetic field applied in a certain direction[2-14]. Such asymmetry is commonly ascribed to some effective field acting at the surface between ferromagnetic and normal layers. As the main influx of information comes from current-voltage measurements, the traditional interpretation involves "spin current" carried by separate electrons and spread due to their physical contact. Transformation of the "spin current" into "charge current" and vice versa is usually ascribed to the spin-orbital coupling manifested itself in form of Rashba effect (inside the hard metal layer) or Dzyaloshinskii-Moria interaction (at the interface). Numerous models involving these two phenomena are developed to explain the observed direction and value of the interface field[11, 15-21]. Moreover, the spin-orbit coupling is considered not only as a main reason for anisotropic magnetoresistance but it is believed responsible for current-induced motion of the domain walls and magnetization switching in the ferromagnetic layer itself[2,4,14,18,20,22-28].

An alternative "magnetic" view consists in action of global magnetic fields instead of specific local exchange interactions. The obvious source is electric current itself producing the circumference magnetic field (often called Oersted field) around the wire. Estimation of this field and comparison with experimentally observed effective values is an important step in interpretation of the obtained data. Though the current-induced magnetic field can be relatively strong (hundreds of Oersteds) at the practically used current density and typical layer thickness, it is usually insufficient to explain the measured value of the effective field generated by the interface[4,9]. Using such estimations, the Oersted field is usually ruled out as a possible contributor to the interface field phenomenon.

However, the direct circumference contribution provided by current is not the only possible source of the global field. The stray field surrounding any ferromagnetic body is potentially a much stronger source. Its theoretical upper limit is $4\pi M$ where $M$ is magnetization. For typical ferromagnets this value is of the order of 10 kOe. The stray field should be especially effective in the perpendicular magnetic anisotropy geometry used in a majority of experiments where the interface field has been reported. Moreover, in case of ferromagnets large change of magnetization (that is, change of configuration and value of stray field) is achieved with moderate values of external field (of the order of coercivity). That is, ferromagnetic layer combined with current can act as a "magnetic amplifier" capable to provide strong output even in the nominally "field-free" experiments[29,30]. In this Letter we shall estimate the stray field acting at the electrons in the adjacent heavy metal layer to produce additional contribution into the Hall effect measured in spintronics.

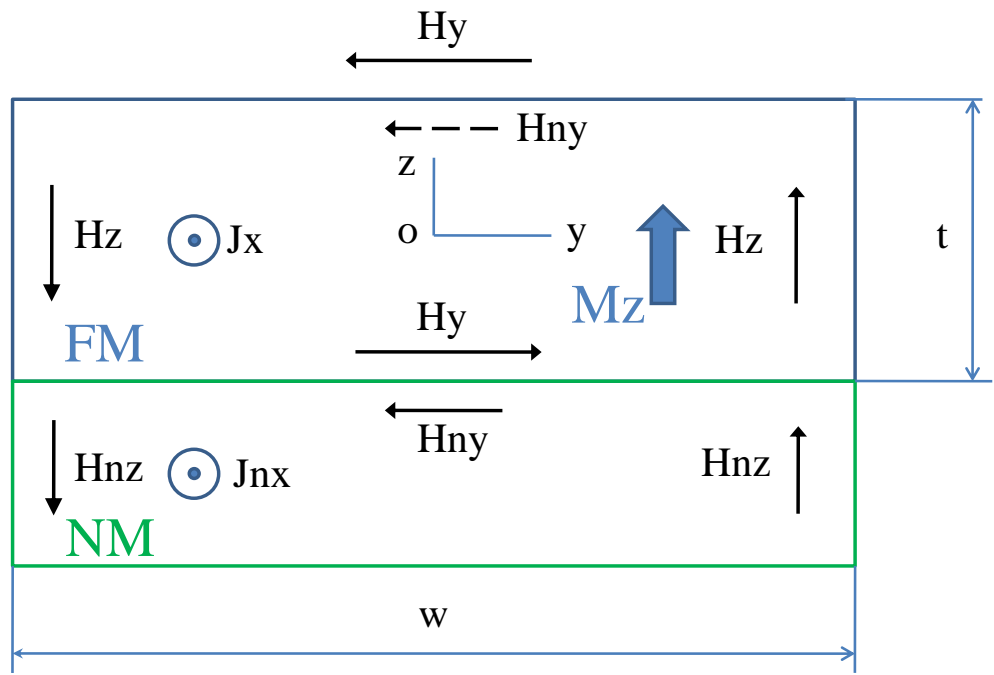

FIG 1. Geometry of the sample, coordinate system and circumference global field due to electric current. FM is a ferromagnetic layer, NM is a normal metal layer.

We consider a flat rectangular ferromagnetic layer of thickness $t$ and width $w$ (Fig. 1). The length of the wire is assumed infinite. The current $J$ with density $j$, $J=jwt$, is directed along the wire length which is chosen as an x axis. This current creates inhomogeneous circumference global magnetic field. Its value can be easily calculated for any point inside and outside the wire (see Refs. 31-33). For our purposes it is important that both components of this field are proportional to the current, so for any important point we can write $H_y=C_y j$ and $H_z=C_z j$ where parameters $C_y$ and $C_z$ are functions of $y$, $z$, $w$ and $t$[31-33]. Here $H_y$ is a component of the global field along the width ($y$ axis) and $H_z$ is directed along the thickness ($z$ axis). Both components are zero in the center of the wire and reach their maximal values at the proper surfaces. Outside the wire the field decreases gradually[31].

Similar field is created by current $j_n$ flowing in the adjacent normal metal layer (Fig. 1). Its components $H_{ny}$ and $H_{nz}$ can be estimated using the same simple formulas as before. Assuming that the field does not decay drastically outside the wire we can see that the components $H_z$ and $H_{nz}$ have the same sign and should be add, while the components $H_y$ and $H_{ny}$ have opposite sign at the interface and identical sign at the free surface of the ferromagnetic layer (Fig. 1).

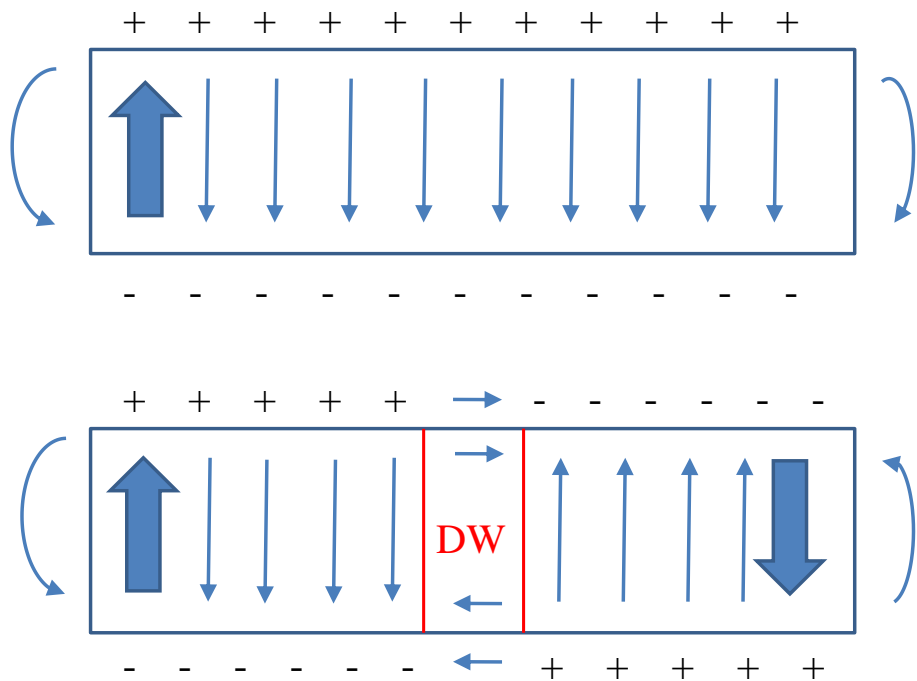

FIG 2. Magnetostatic stray field (thin arrows) due to surface magnetic charges. Top: homogeneous magnetization (thick arrow), bottom: two domains and domain wall (DW).

Now let us consider the stray field induced by magnetization. Most experiments deal with perpendicular magnetic anisotropy so that initial magnetization $M_z$ points along the $z$ axis (Figs. 1, 2). The homogeneously magnetized layer can be approximated with two close surfaces bearing opposite "magnetic charges" with surface density ±*M*. Far from the edges the fields created by these two sources are completely subtracted, so the field is contained inside the sample and near its edges, similar to electric field of the flat capacitor (Fig. 2, top). However, the situation is completely different when the sample contains domain walls (Fig. 2, bottom). Surface magnetic charges around the wall have opposite signs, so the stray field $H_d$ has a strong horizontal component near the surface. For moderate film thickness this field deforms the wall structure inducing the "twisted domain walls" widely discussed in the past[34-36]. In case of the extremely thin films the twisting of walls should be damped by fast growth of the exchange energy; nevertheless the strong horizontal stray field still exists at both sides of the surface. To estimate it most researchers use the expression given by Hagedorn[35]

$$H_{dy} = 2M \left[ \ln \frac{4t^2+\delta^2}{\delta^2} + 4 \frac{t}{\delta} \tan^{-1} \frac{\delta}{2t} \right] \qquad (1)$$

Here δ is effective width of the domain wall and *M* is magnetization. This approximation has been criticized by O'Dell[36] who proposed

$$H_{dy} = 2M \qquad (2)$$

as a more realistic estimation. Even the smaller of these two values still gives $H_d$~1600 Oe for most of typical ferromagnets.

Now we can compare these purely magnetic contributions with experimentally observed values of the effective interface field measured in spintronic experiments to decide whether they can provide an alternative explanation to commonly assumed spin-orbit mechanisms. There is one qualitative disagreement to explain in this way. The stray field at the domain wall has fixed value and direction. Meanwhile, the interface field measured in the reported experiments[4,5,7,9,10,12] has different components and often seems to be proportional to current. The stray field is strong but fixed, while both components of the Oersted field vary with current but they are too weak. To resolve this discrepancy we have to take into account tilting of the wall by vertical components of the global field at the lateral edges (Fig. 3). Detailed consideration of this phenomenon can be found, for example, in Ref. 37. The vertical component $H_z$ of the global field has opposite sign at the edges, so it pushes the domain wall in opposite directions. For small values of current the wall simply turns around the *z* axis, then its shape becomes nonlinear, and at last the wall turns parallel to the x axis above some critical current[37]. The stray field at the surface is perpendicular to the wall and for *j*>0 it acquires a component $H_{dy} = H_d \sin \theta = H_d j/j_c$ where $j_c$ is critical current density calculated in Ref. 37. Now we can write all three components of the magnetic field acting at the interface:

$$H_x = 2M \cos \theta \qquad (3)$$

$$H_y = \left(2M/j_c + (C_y - C_{yn})\right) j \qquad (4)$$

$$H_z = (C_z + C_{zn})j + H_{dz} \qquad (5)$$

The last term in Eq. 5 describes vertical component of stray field near the edges of the ferromagnetic layer (Fig. 2). We see that the two components of magnetic field, $H_y$ and $H_z$, vary linearly with current, while dependence of $H_x(j)$ is nonlinear and should be considered separately.

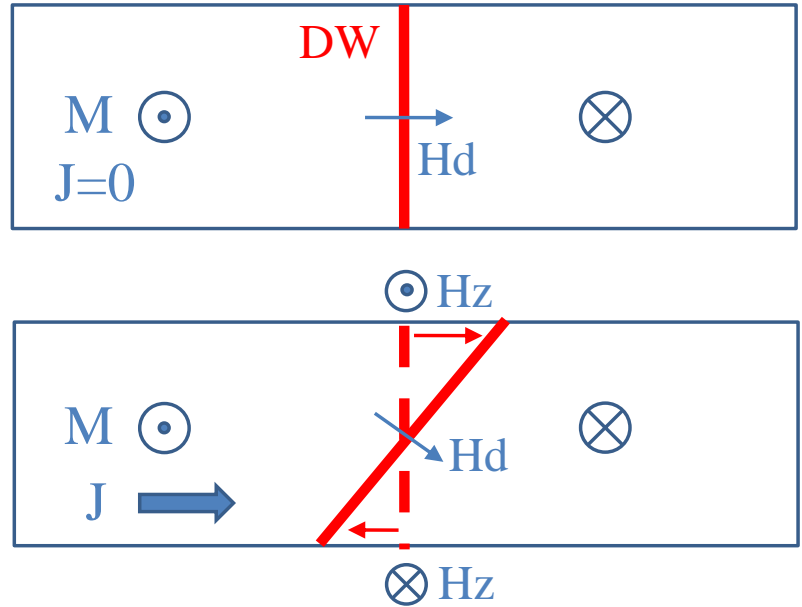

FIG3. Orientation of the domain wall and stray field at the surface (top view). Top: initial domain wall is parallel to the *y* axis. Bottom: in presence of current the wall is tilted due to pressure by vertical component of the global field, and the stray field acquires the *y* component.

Using for numerical estimations $M=1423$ G (cobalt, the most widely used material in experiments with perpendicular magnetic anisotropy) and $\theta=20°$ we have $H_x=2674$ Oe and $H_y=968$ Oe even neglecting Oersted field which is relatively small. According to results of Ref. 37, such angle should be achieved with current density of the order of $j=0.34\ j_c$ where $j_c \sim 2.2 \times 10^{12}$ A/m$^2$ for typical experiments[37]. Expected ratio of effective field to current is $H_y/j=588$ Oe/$10^{12}$ A/m$^2$. The experimentally measured values are scattered from 100 to 800 Oe/$10^{12}$ A/m$^2$ (Refs. 5,7,9,10,12) with most typical result of 500..600 Oe/$10^{12}$ A/m$^2$ (Refs. 5,6). We have to conclude that our rough estimation gives the magnetostatic interface field very close to experimental data.

The most common explanations of the interface field concern spin-orbit interaction in the heavy metal layer[4,18,22,26,38]. The value and direction of the surface field are discussed in terms of equations like

$$\boldsymbol{H_S} = R\ \boldsymbol{z} \times (\boldsymbol{M} \times \boldsymbol{j}) \tag{6}$$

Here $z$ is unit vector along the interface normal, R is a proper constant[2,18]. In our coordinate system $j$ is directed along $x$ and $M$ is directed along $z$, so the main component of $H_S$ is directed along $y$ axis and changes its sign with reversal of current. It is easy to see from Eq. 4 that the $y$ component of the magnetostatic stray field has similar features. This means that measuring of the interface field itself is not sufficient to decide whether it is of magnetostatic or spin-orbit origin, so special validation is required in any case to distinct between these mechanisms.

We have to mention also that the considered distribution of field is not symmetric across the ferromagnetic layer. The value of magnetic field at the two surfaces is different due to the field produced by the current in the normal metal layer. The two components $H_y$ and $H_{ny}$ add at one surface and cancel at the other one (Fig. 1). This means that energy of the domain wall can be different at both surfaces if the magnetization in the wall has a Bloch component (directed along or opposite to this field). The low thickness of the film (below the exchange length) prevents change of magnetization across the film thickness but this is not the case for magnetic field which can vary very sharply. Frustration of the wall configuration should result to some intermediate structure corresponding to the minimum of the total energy. Consideration of this problem needs additional assumptions on the wall structure and is not discussed in this Letter; however, the picture seems potentially rich with additional mechanisms of the preferred wall chirality and current-induced creep (for example, due to periodic change of the tilting angle θ).

In conclusion, we have estimated direction and value of magnetostatic stray field in the ferromagnetic layer with perpendicular magnetic anisotropy and its potential contribution to the interface field measured in spintronic experiments. Due to tilting of domain walls caused by the vertical part of current-induced global field the stray field acquires additional transverse component proportional to current. The estimated value and direction of the field-current ratio are in good agreement with reported experimental data. The proposed mechanism should be taken into account considering current-induced domain wall motion and spin-Hall experiments.

**References**

[1] S.S.P. Parkin, M. Hayashi, and L. Thomas, Science **320**, 190 (2008).

[2] I.M. Miron, K. Garello, G. Gaudin, P.-J. Zermatten, M.V. Costache, S. Auffret, S. Bandiera, B. Rodmacq, A. Schuhl, and P. Gambardella, Nature **476**, 189 (2011).

[3] T. Suzuki, S. Fukama, N. Ishiwata, M. Yamanouchi, S. Ikeda, N. Kasai, and H. Ohno, Appl. Phys. Lett. **98**, 142505 (2011).

[4] L. Liu, O.J. Lee, T.J. Gudmundsen, D.C. Ralph, and R.A. Buhrman, Phys. Rev. Lett. **109**, 096602 (2012).

[5] S. Emori, D.C. Bono, and G.S.D. Beach, Appl. Phys. Lett. **101**, 042405 (2012).

[6] S. Emori, U. Bauer, S.-M. Ahn, E. Martinez, and G.S.D. Beach, Nature Mater. **12**, 611 (2013).

[7] M. Kawaguchi, K. Shimamura, S. Fukami, F. Matsukura, H. Ohno, T. Moriyama, D. Chiba, and T. Ono, Appl. Phys. Express **6**, 113002 (2013).

[8] C. Zhang, M. Yamanouchi, H. Sato, S. Fukami, S. Ikeda, F. Matsukura, and H. Ohno, Appl. Phys. Lett. **103**, 262407 (2013).

[9] H.-R. Lee, K. Lee, J. Cho, Y.-H. Choi, C.-Y. You, M.-H. Jung, F. Bonell, Y. Shiota, S. Miwa, andY. Suzuki, Sci. Rep. **4**, 6548 (2014).

[10] S. Woo, M. Mann, A.J. Tan, L. Caretta, and G.S.D. Beach, Appl. Phys. Lett. **105**, 212404 (2014).

[11] P. Li, T. Liu, H. Chang, A. Kalitsov, W. Zhang, G. Czaba, W. Li, D. Richardson, A. DeMann, G. Rimal, H. Dey, J.S. Jiang, W. Porod, S.B. Field, J. Tang, M.C. Marconi, A. Hoffmann, O. Mryasov, and M. Wu, Nature Comm. **7**, 12688 (2016).

[12] J. Li, G. Yu, Y. Liu, Z. Shi, Y. Liu, A. Navabi, M. Aldosary, Q. Shao, K.L. Wang, R. Lake, and J. Shi, Phys. Rev. B **95**, 241305 (2017).

[13] N. Kato, M. Kawaguchi, Y.-C. Lau, T. Kikuchi, Y. Nakatani, and M. Hayashi, Phys. Rev. Lett. **122**, 257205 (2019).

[14] M. Yamanouchi, N.V. Bao, M. Inoue, and T. Uemura, Jpn. J. Appl. Phys. **58**, 100903 (2019).

[15] M.D. Stiles and A. Zangwill, Phys. Rev. B **66**, 014407 (2002)

[16] S. Urazhdin, Phys. Rev. B **78**, 060405 (2008).

[17] P.M. Haney, H.-W. Lee, K.-J. Lee, A. Manchon, and M.D. Stiles, Phys. Rev. B **88**, 214417 (2013).

[18] E. Martinez, S. Emori, and G.S.D. Beach, Appl. Phys. Lett. **103**, 072406 (2013).

[19] I.A. Ado, O.A. Tretiakov, and M. Titov, Phys. Rev. B **95**, 094401 (2017).

[20] M. Li, J. Wang, and J. Lu, New J. Phys. **21**, 053011 (2019).

[21] J. Ryu, S. Lee, K.-J. Lee, and B.-G. Park, Adv. Mater. **32**, 1907148 (2020).

[22] J. Linder, Phys. Rev. B **87**, 054434 (2013).

[23] E. Martinez, S. Emori, N. Perez, L. Torres, and G.S.D. Beach, J. Appl. Phys. **115**, 213909 (2014).

[24] Y. Kurokawa, M. Kawamoto, and H. Amano, Jpn. J. Appl. Phys. **55**, 07MC02 (2016).

[25] J. Yu, X. Qiu, Y. Wu, J. Yoon, P. Deorani, J.M. Besbas, A. Manchon, and H. Yang, Sci. Rep. **6**, 32629 (2016).

[26] K.-J. Kim, Y. Yoshimura, and T. Ono, Jpn. J. Appl. Phys. **56**, 0802A4 (2017).

[27] T. Taniguchi, J. Grollier, and M.D. Stiles, Phys. Rev. Applied **3**, 044001 (2015).

[28] H. Wu, S.A. Razawi, Q. Shao, X. Li, K.L. Wong, Y. Liu, G. Yin, and K.L. Wang, Phys. Rev. B **99**, 184403 (2019).

[29] I. Gilbert, P.J. Chen, D.B. Gopman, A.L. Balk, D.T. Pierce, M.D. Stiles, and J. Unguris, Phys. Rev. B **94**, 094429 (2016).

[30] M. Chang, J. Yun, Y. Zhai, B. Cui, Y. Zuo, G. Yu, and L. Xi, Appl. Phys. Lett. **117**, 142404 (2020).

[31] R. Berthe, A. Birkner, and U. Hartmann, Phys. Stat. Sol. (a) **103**, 557 (1987).

[32] N. Smith, W. Doyle, D. Markham, and D. LaTourette, IEEE Trans. Magn. **23**, 3248 (1987).

[33] S. LeGall, F. Montaigne, D. Lacour, M. Hehn, N. Vernier, D. Ravelosona, S. Mangin, S. Andrieu, and T. Hauet, Phys. Rev. B **98**, 024401 (2018).

[34] J.C. Slonczewski, J. Appl. Phys. **44**, 1759 (1973).

[35] F.B. Hagedorn, J. Appl. Phys. **45**, 3129 (1974).

[36] T.H. O'Dell, Phys. Stat. Sol. (a) **48**, 59 (1978).

[37] S.-C. Yoo, K.-W. Moon, and S.-B. Choe, J. Magn. Magn. Mater. **343**, 234 (2013).

[38] A. Manchon and S. Zhang, Phys. Rev. B **79**, 094422 (2009).